\documentclass[iop,apjl]{emulateapj}

\shorttitle{A resolved debris disk around HD95086}

\shortauthors{Mo\'or et al.}

\begin{document}


\title{A resolved debris disk around the candidate planet-hosting star HD95086}

\author{A. Mo\'or\altaffilmark{1}}
\email{moor@konkoly.hu}
\author{P. \'Abrah\'am\altaffilmark{1}}
\author{\'A. K\'osp\'al\altaffilmark{2}}
\author{Gy.~M. Szab\'o\altaffilmark{1,3,4}}
\author{D. Apai\altaffilmark{5}}
\author{Z. Balog\altaffilmark{6}}
\author{T. Csengeri\altaffilmark{7}}
\author{C. Grady\altaffilmark{8,9}} 
\author{Th.~Henning\altaffilmark{6}}
\author{A. Juh\'asz\altaffilmark{10}}
\author{Cs. Kiss\altaffilmark{1}}
\author{I. Pascucci\altaffilmark{5}}
\author{J. Szul\'agyi\altaffilmark{11}}
\author{R. Vavrek\altaffilmark{12}}

\altaffiltext{1}{Konkoly Observatory, Research Centre for Astronomy
  and Earth Sciences, Hungarian Academy of Sciences, PO Box 67, H-1525
  Budapest, Hungary}

\altaffiltext{2}{European Space Agency (ESA/ESTEC, SRE-SA), P.O. Box
  299, 2200 AG, Noordwijk, The Netherlands ; ESA fellow}

\altaffiltext{3}{ELTE Gothard Astrophysical Observatory, Szent Imre herceg \'ut 112, H-9700 Szombathely, Hungary}

\altaffiltext{4}{Dept. of Experimental Physics and Astronomical
  Observatory, 6720 Szeged D\'om t\'er 9., Hungary}

\altaffiltext{5}{Department of Astronomy and Department of Planetary
  Sciences, The University of Arizona, Tucson, AZ 85721}

\altaffiltext{6}{Max-Planck-Institut f\"ur Astronomie, K\"onigstuhl
  17, 69117 Heidelberg, Germany}

\altaffiltext{7}{Max-Planck-Institut f\"ur Radioastronomie, Auf dem
  H\"ugel 69, 53121 Bonn, Germany}

\altaffiltext{8}{NASA Goddard Space Flight Center, Code 667,
  Greenbelt, MD 20771, USA}

\altaffiltext{9}{Eureka Scientific, 2452 Delmer Street, Suite 100,
  Oakland, CA 94602, USA}

\altaffiltext{10}{Leiden Observatory, Leiden University, Niels Bohrweg 2, NL-2333 CA Leiden, The Netherlands}

\altaffiltext{11}{Universit\'e de Nice Sophia-Antipolis, Observatoire
  de la C\^ote d'Azur, CNRS UMR 7293, 06108, Nice Cedex 2, France}

\altaffiltext{12}{Herschel Science Centre, ESA/ESAC, PO Box 78,
  Villanueva de la Ca\~nada, 28691, Madrid, Spain}


\begin{abstract}

Recently, a new planet candidate was discovered on direct images around the young (10-17 Myr) A-type
star HD95086. The strong infrared excess of the system indicates that, similarly
to HR8799, $\beta$~Pic, and Fomalhaut, the star harbors a circumstellar disk. Aiming to study the
structure and gas content of the HD95086 disk, and to investigate its possible interaction with the
newly discovered planet, here we present new optical, infrared and millimeter observations. 
We detected no CO emission, excluding the possibility of an evolved gaseous primordial disk.
Simple blackbody modeling of the spectral energy distribution suggests the presence of two spatially separate dust
belts at radial distances of 6 and 64~AU. 
Our resolved images obtained with the {\sl Herschel Space Observatory} 
reveal a characteristic disk size of
$\sim6\farcs0\times5\farcs4$ (540$\times$490\,AU) and disk inclination of $\sim$25$^{\circ}$. 
Assuming the same inclination
for the planet candidate's orbit, its re-projected radial distance from the star is 62~AU, very close
to the blackbody radius of the outer cold dust ring.
The structure of the planetary system at HD95086 resembles 
the one around HR8799. Both systems harbor a warm inner dust belt
and a broad colder outer disk and giant planet(s) between the two
dusty regions. Modelling implies that the
candidate planet can dynamically excite the motion of planetesimals
even out to 270~AU via their secular perturbation if its orbital
eccentricity is larger than about 0.4.
Our analysis adds a new
example to the three known systems where directly imaged planet(s) and debris disks co-exist.

\end{abstract}






\section{Introduction}
\label{intro}

Recently, several new massive exoplanets at large orbital radii were
directly imaged, e.g.~around HR8799 \citep{marois2008}, $\beta$~Pic
\citep{lagrange2009}, and Fomalhaut \citep{kalas2008}. They are the
largest products of the planet formation process, which also produces
many smaller planetesimals. While these smaller bodies cannot directly be
detected, {\it debris} dust arising from their erosion can be
observed via scattered light and/or thermal emission of the
grains. All three mentioned examples harbor luminous debris
disks. Giant planet(s) and the planetesimal belt/debris disk can interact with each other   
in several ways.
A giant planet can sculpt the structure of the
debris disk \citep{ertel2012},  
while the planetesimal disk can also influence the planet's
orbital evolution and long-term stability \citep{moore2013}.

\citet{rameau2013} discovered a new planet candidate around HD95086, 
at a projected separation of 56~AU. Its estimated
mass of 4--5 M$_{\rm Jup}$ makes it the lowest mass planet detected by
direct imaging. The host star, similarly to HR8799 and $\beta$ Pic,
is an A-type star, belongs to a young association \citep[Lower
  Centaurus Crux,][hereafter, LCC]{dezeeuw1999}, and exhibits
prominent infrared excess indicative of a dusty circumstellar disk
\citep{rhee2007,chen2012,rizzuto2012}. The age estimates of the
LCC association range between 10~Myr \citep{song2012} and 17~Myr \citep{pecaut2012}.
While the disk
of HD95086 probably contains secondary dust, {considering} its young age, {it is 
possible} that the disk {is} an evolved
gaseous primordial disk.
Here, we study the structure and gas
content of the HD95086 disk and investigate its possible
interaction with the newly discovered planet.
Our analysis adds a new example to the three known 
systems where directly imaged planet(s) and debris disks co-exist.

\section{Observations and data reduction} \label{obsanddatared}

We observed HD95086 with the {\sl Herschel Space Observatory}
\citep{pilbratt2010} using the Photodetector Array Camera and
Spectrometer \citep[PACS,][]{poglitsch2010}, and the Spectral and
Photometric Imaging Receiver \citep[SPIRE,][]{griffin2010}. PACS
observations were performed on 2011 August 1 in mini scan map mode with
medium scan speed (20{\arcsec}~s$^{-1}$). {Measurements at scan angles of
70{$\degr$} and 110{$\degr$} were made both at 70~{\micron} and
100~{\micron} and this setup also provided four observations at 160{\micron}.
Each measurement included four repetitive scan maps.}  
Data pocessing was carried
out with the Herschel Interactive Processing Environment
\citep[HIPE,][]{ott2010}, v.9.2, using the pipeline optimized for
bright sources. 1/f noise was removed by highpass filtering, after our
target was masked to avoid flux loss. We used second-level
deglitching to remove glitches. Mosaics (Fig.~\ref{fig1}) were created
with pixel sizes of 1$\farcs$1, 1$\farcs$4, and 2$\farcs$1 at 70, 100,
and 160~{\micron}, by combining the individual scan maps with a
weighted average. SPIRE maps were obtained on 2011 August 16 at 250,
350 and 500~{\micron} {in} small scan map mode with a
repetition factor of 2. Data reduction was performed with HIPE v9.2
using the standard pipeline script. The beam size was 18$\farcs$1,
25$\farcs$2 and 36$\farcs$6 at 250, 350, and 500~{\micron},
respectively, and the maps were resampled to pixel sizes of
6{\arcsec}, 10{\arcsec}, and 14{\arcsec}.

HD95086 was observed with the InfraRed Spectrograph (IRS) onboard the
{\sl Spitzer Space Telescope} on 2004 February 4. Small 2$\times$3 maps were
taken using the low-resolution IRS modules, covering the
$5.2-38\,\mu$m wavelength range with a spectral resolution of
$R=60-120$. We downloaded the data processed with the pipeline version
S18.18.0 from the archive and further processed them with the
Spitzer IRS Custom Extraction Software (SPICE v2.5.0). We treated the
data as a nodding measurement by using the two central map positions
and subtracting them from each other. Then, we extracted the positive
signal from a wavelength-dependent, tapered aperture, and averaged
them. The final spectrum is plotted in Fig.~\ref{fig2}a.

On 2011 April 22, we observed HD95086 at the 345.796\,GHz $^{12}$CO
$J$=3$-$2 line using the SHeFI/APEX2 receiver \citep{vassilev2008}
mounted at the 12~m APEX telescope\footnote{This publication is based
  on data acquired with the Atacama Pathfinder EXperiment (APEX). APEX
  is a collaboration between the Max-Planck-Institut f\"ur
  Radioastronomie, the ESO, and the Onsala
  Space Observatory.} \citep{gusten2006} (M-087.F-0001 program, PI:
Th.~Henning). We used the Fast Fourier Transform Spectrometer with
2048 channels, providing a velocity resolution of 0.42\,km\,s$^{-1}$.
An on-off observing pattern was utilized with beam switching. The
total on-source integration time was 29.4\,minutes. The data reduction
was performed using
GILDAS/CLASS\footnote{\url{http://iram.fr/IRAMFR/GILDAS/}}. For the
final spectrum, we subtracted a linear baseline from each individual
scan, and averaged them after omitting the noisy scans.

We obtained a high-resolution optical spectrum of HD95086 with the
Fiber-fed Extended Range Optical Spectrograph
\citep[FEROS,][]{kaufer1999} mounted at the 2.2\,m MPG/ESO telescope
in La Silla, Chile, on 2011 April 17. This instrument covers the
wavelength range between 3500 and 9200 \AA{} in 39 echelle orders with
a {resolution} of $R\approx$48~000. We used the ``object-sky''
mode, with one fiber positioned at the target, and the other one on
the sky. The integration time was 120~s. Data reduction, including
bias subtraction, flat-field correction, background subtraction, the
definition and extraction of orders, and wavelength calibration, was
performed using the FEROS data reduction system pipeline at the
telescope.

\section{Results and analysis}
\label{analysis}

\subsection{Stellar properties}
\label{stellarprops}

To estimate the basic properties of HD95086 and provide
photospheric flux predictions at relevant mid- and far-IR wavelengths,
we modeled the stellar photosphere by fitting the optical and infrared
photometric data with an ATLAS9 atmosphere model \citep{castelli2003}.
We used photometry from {\sl Hipparcos} \citep{perryman1997}, {\sl
  TYCHO2} \citep{hog2000}, and 2MASS \citep{skrutskie} that were
supplemented by $W1$ band (centered on 3.4{\micron})
photometry from {\sl WISE} \citep{wright2010}. Assuming that the
interstellar reddening is negligible and adopting solar metallicity
and $\log{g} = 4.0,$ our $\chi^2$ minimization yield{ed} $T_{\rm eff}$ =
7500$\pm$150\,K and a luminosity of $L_{\rm bol}$ = 7.0$\pm$0.6
L$_\odot$ (using a ${\sl Hipparcos}$-based distance of 90.4\,pc,
\citealt{vanleeuwen2007}). The derived $T_{\rm eff}$ is identical with {that}
obtained by \citet{chen2012} from color indices.

To obtain $T_{\rm eff}$, $\log\ g$, and $v\sin\ i$ from our FEROS
spectrum, we used an iterative fitting method described in
e.g. \citet{szabo2011}. First, 
we varied these three parameters using
spectra from \citet{munari2005}, in the
4000--6200\AA\ wavelength range, excluding the $H_\beta$ region and
Na~D lines. Radial velocity was calculated from the cross-correlation
function, convolving the measured spectrum with the best-fit Munari
template, but setting $\log\ g$ to 1.0 and $v\sin\ i$ to 0. This
yielded 17$\pm 2$\,km\,s$^{-1}$. After transforming the measured
spectrum to the laboratory system, and keeping $T_{\rm eff}$,
$\log\ g$, and $v\sin\ i$ in the global minimum, we fitted [Fe/H] in
the second step. $T_{\rm eff}$ and $\log\ g$ was re-fitted in the
third iteration. Our best-fit parameters are $T_{\rm eff} =
7750\pm250$K, $\log\ g = 4.0\pm0.5$, $v\sin\ i =
20\pm10$kms$^{-1}$, [Fe/H] = $-$0.25$\pm$0.5. 
This $T_{\rm eff}$ supports our result obtained from the photometry 
(see above). 
\citet{rameau2013} estimated a stellar mass of 1.6~M$_{\odot}$ for HD95086.

\subsection{Herschel maps} \label{pacsanalysis}

\begin{figure*} 
\includegraphics[scale=.30,angle=0]{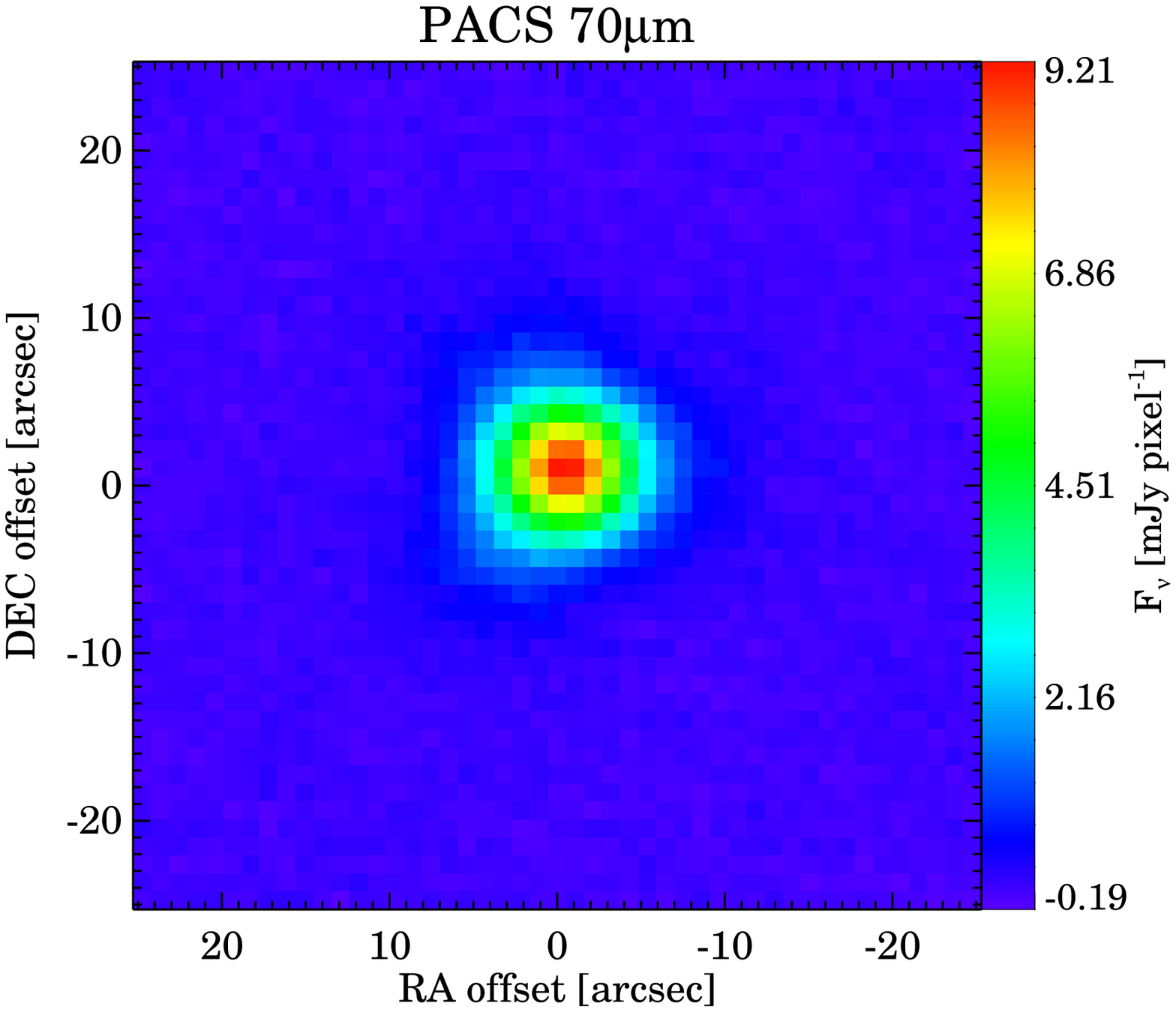}
\includegraphics[scale=.30,angle=0]{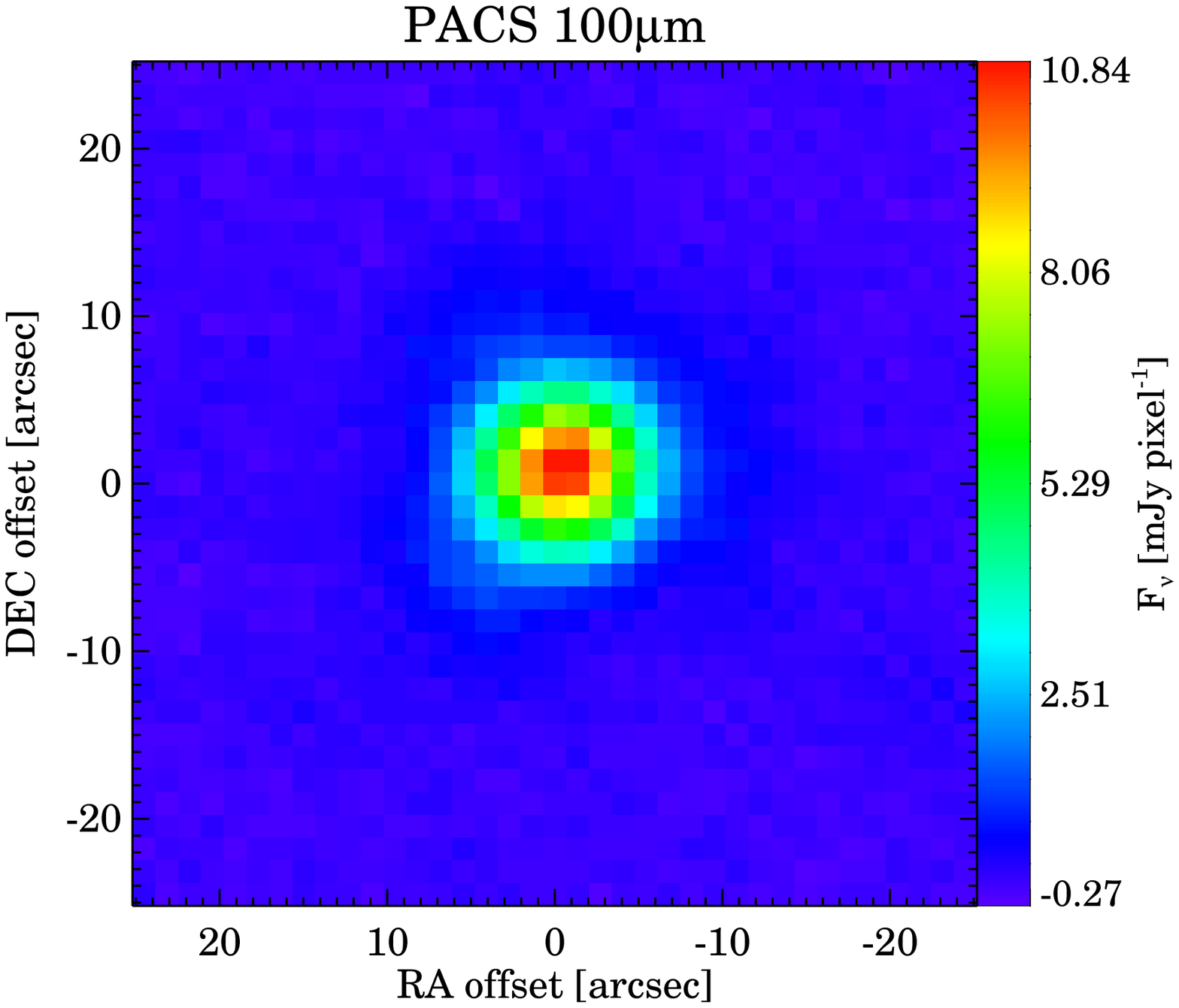}
\includegraphics[scale=.30,angle=0]{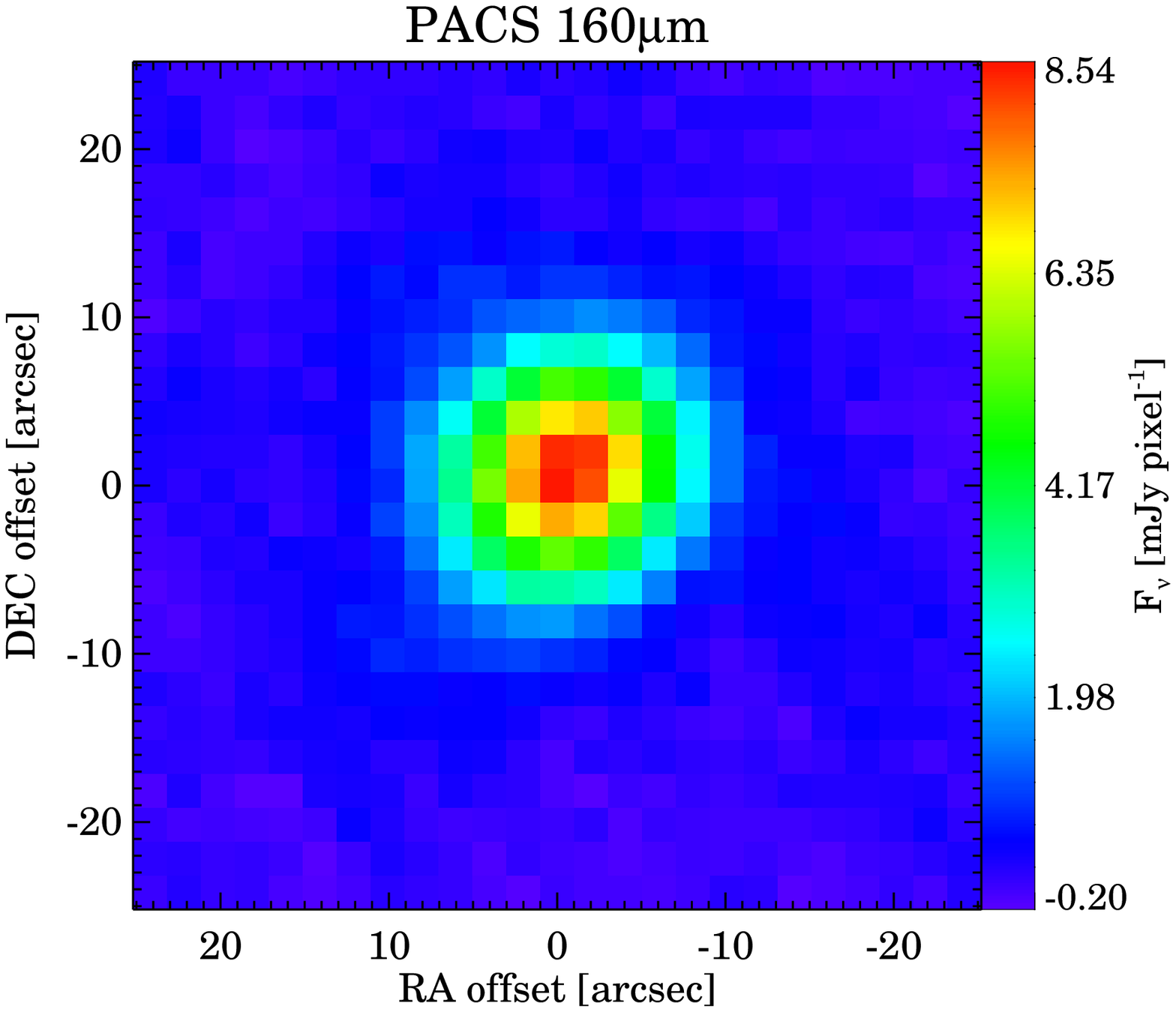} \\
\includegraphics[scale=.30,angle=0]{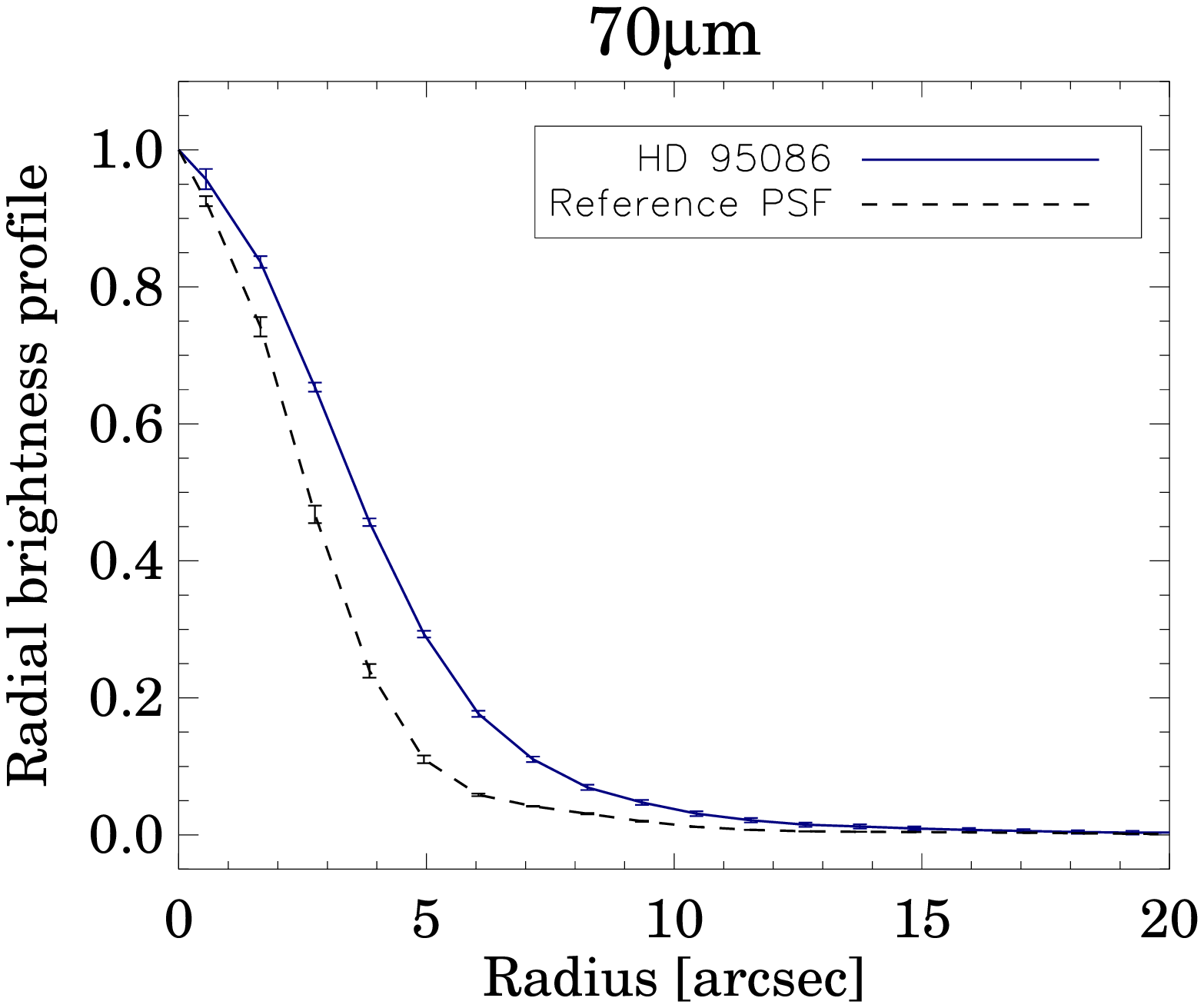}
\includegraphics[scale=.30,angle=0]{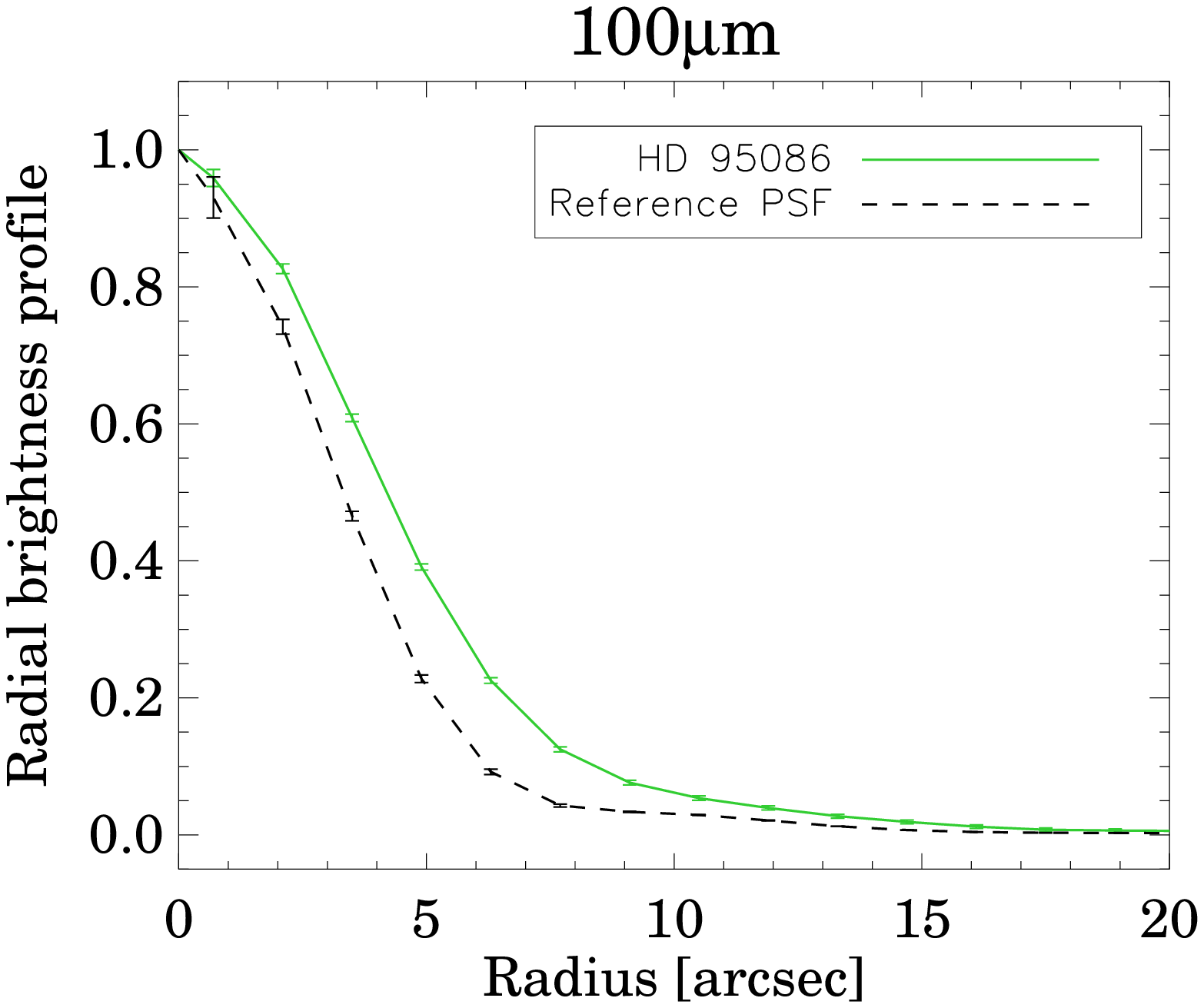}
\includegraphics[scale=.30,angle=0]{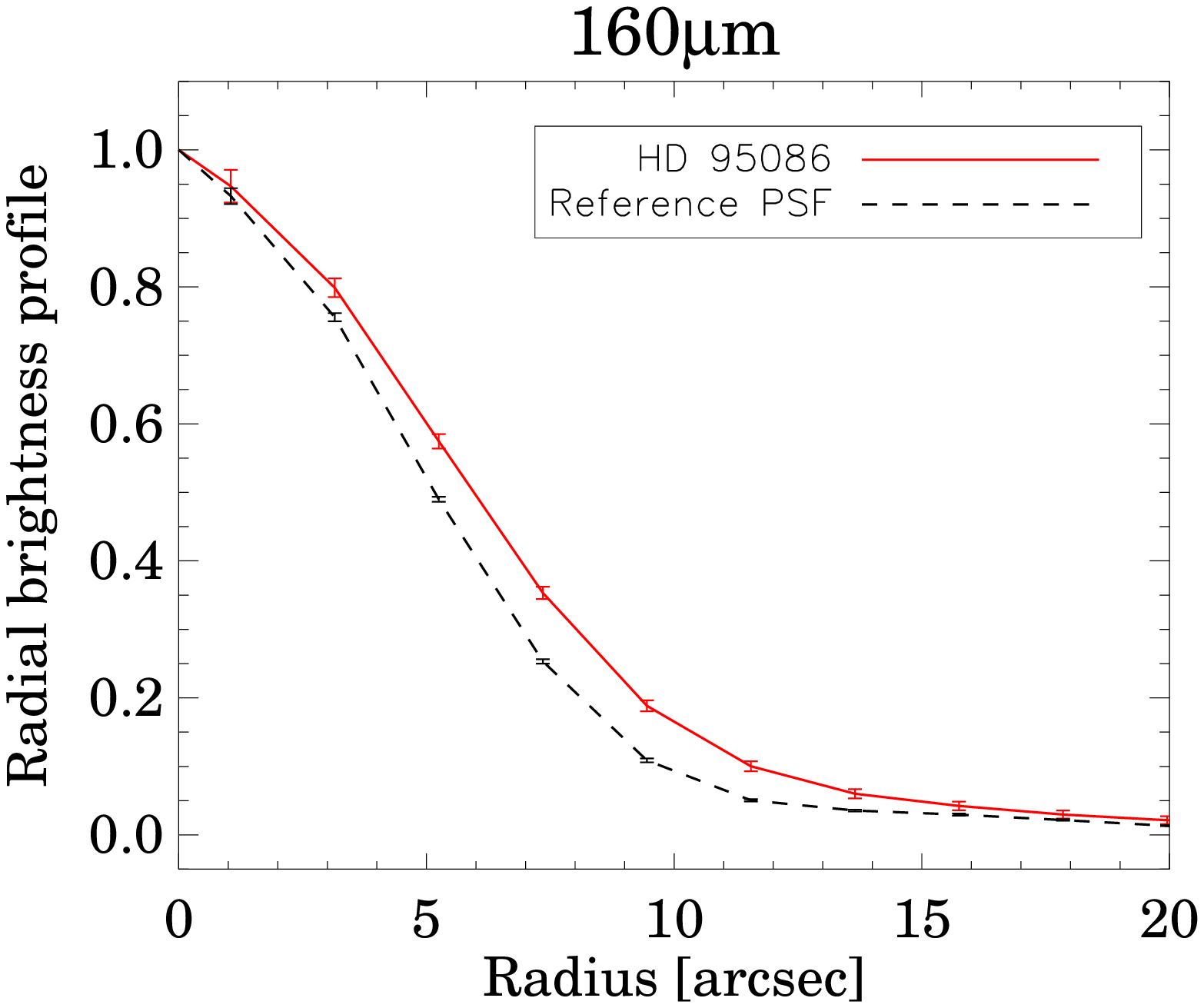}
\caption{ {\sl Upper: Herschel PACS 70/100/160{\micron} images of HD95086. Lower: 
radial brightness profiles for HD95086 and reference PSF profiles at the same wavelengths 
(see Sect.~\ref{pacsanalysis}). }
\label{fig1}
}
\end{figure*}

HD95086 appeared as an isolated bright source on our PACS images
(Fig.~\ref{fig1}, upper panels). To estimate the disk size,
we constructed point spread functions (PSFs) using PACS observations
of two calibrator stars ($\alpha$~Boo, $\alpha$~Tau) without known
infrared excess, measured and processed in the same way as HD95086.
The PSF maps were rotated to match the roll angle of the telescope at
the time of observing HD95086. 
{The FWHM point-source sizes, as measured on these maps are 5$\farcs$5$\pm$0$\farcs$1, 
6$\farcs$7$\pm$0$\farcs$1, and 10$\farcs$6$\pm$0$\farcs$3 at
70, 100, and 160{\micron}, respectively.} 
We derived azimuthally averaged radial
brightness profiles and averaged the individual measurements to obtain
a reference PSF. By comparing the azimuthally averaged radial
brightness profiles of HD95086 with these reference PSFs, we
found that the disk around HD95086 is spatially extended at all PACS
wavelengths (Fig.~\ref{fig1}, lower panels).

{We fitted elliptical Gaussians to the images of the source in order to derive the FWHM sizes of 
the major and minor axes and the position angles (see Table~\ref{tabdeconv}).}   
{ 
Then, we estimated the disk size by quadratic deconvolution with the PSF, and calculated the inclination 
(at 70 and 100{\micron})
assuming that
the intrinsic shape/structure of the disk is azimuthally symmetric.
} 
The contribution of stellar photosphere to the total flux
was negligible ($<$0.3\%) in all bands, thus we did not subtract the
model photospheric emissions from the source images before the
Gaussian fitting.
Table~\ref{tabdeconv} lists the final parameters and their uncertainties.
 The coarse spatial
resolution at 160~{\micron} prevented us from determining a {reliable} position
angle and inclination. Our results outline a large disk with a 
characteristic size of $\sim$6\farcs0$\times$5\farcs4 ($\sim$540$\times$490~AU).    

\begin{deluxetable*}{lccccc}
\tabletypesize{\scriptsize}
\tablecaption{Disk properties derived from quadratic deconvolution (see Sect~\ref{pacsanalysis}) \label{tabdeconv}}
\tablewidth{0pt}                                                                                                                                                                    
\tablecolumns{6}
\tablehead{ \colhead{} & \colhead{Original FWHM of the} & \multicolumn{2}{c}{Disk size after}  & \colhead{Position} & \colhead{Incl.} \\
\colhead{} & \colhead{source before deconvolution} & \multicolumn{2}{c}{quadratic deconvolution}  & \colhead{angle} & \colhead{} \\
\colhead{}  & \colhead{major$\times$minor axis [{\arcsec}]} &  \colhead{major$\times$minor axis [{\arcsec}]} &  \colhead{major$\times$minor axis [AU]} & [{$\degr$}] & [{$\degr$}] 
}
\startdata  
PACS 70{\micron}   & 8.1$\pm$0.1$\times$7.7$\pm$0.1 & 5.9$\pm$0.1$\times$5.4$\pm$0.2 & $\sim$530$\times$490 &  99$\pm$5  & 23$\pm$5   \\
PACS 100{\micron}  & 9.1$\pm$0.1$\times$8.6$\pm$0.1 & 6.2$\pm$0.2$\times$5.4$\pm$0.2 & $\sim$560$\times$490 & 113$\pm$7  & 29$\pm$4   \\
PACS 160{\micron}  & 12.5$\pm$0.5$\times$12.2$\pm$0.4 & 6.7$\pm$1.0$\times$6.1$\pm$0.9 & $\sim$600$\times$550 &   \ldots       &  \ldots  \\
\enddata                                                                        
\end{deluxetable*} 

We measured the flux of HD95086 on the individual PACS scan maps
using an aperture radius of 18$''$ and a sky annulus between 40$''$
and 50$''$. We calculated the average and RMS of the individual
flux values, applied aperture correction, and calculated the final
uncertainties of the photometry by adding quadratically the
measurement errors and an absolute calibration uncertainty of 7\%
(Balog et al., submitted). Our target was clearly detected as a point
source at all SPIRE wavelengths. We performed photometry with an
aperture radius of 22{\arcsec}, 30{\arcsec}, and 42{\arcsec} at 250,
350, and 500~{\micron}. Background levels were estimated in annuli
extending from 60{\arcsec} to 90{\arcsec}. The final uncertainties
were derived as the quadratic sum of the measurement errors and the
overall calibration uncertainty of 5.5\% for the SPIRE
photometer \citep{bendo2013}. Our Herschel
photometry for HD95086 is listed in Table~\ref{phottable}.

\subsection{Modeling of dust distribution}
\label{modelling}

We compiled the spectral energy distribution (SED) of HD95086 by
combining the new PACS and SPIRE fluxes with infrared photometry from
the literature. For the fitting process the IRS spectrum was sampled
in 11 bins. Table~\ref{phottable} summarizes the collected IR
data. 
\begin{deluxetable*}{ccccc}                                                      
\tabletypesize{\scriptsize}                                                     
\tablecaption{Measured and predicted fluxes \label{phottable}}                  
                                                                                
\tablewidth{0pt}                                                                
                                                                                
\tablecolumns{5}                                                                
                                                                                
\tablehead{ \colhead{Wavelength} & \colhead{Measured flux density} &                    
\colhead{Instrument} &  \colhead{Predicted photospheric flux density} &  \colhead{Reference} \\      
\colhead{[{\micron}]} & \colhead{[mJy]} &                                       
\colhead{} &  \colhead{[mJy]} &  \colhead{}                                     
}                                                                               
\startdata                                                                      
     3.35 &         626.7$\pm$24.7 &               WISE &    592.3 &   \citet{wright2010} \\
     4.60 &         336.8$\pm$11.5 &               WISE &    329.4 &   \citet{wright2010} \\
     9.03 &          95.7$\pm$13.0 &                IRS &     90.7 &            this work \\
    11.02 &           67.3$\pm$7.0 &                IRS &     61.3 &            this work \\
    11.56 &           62.7$\pm$3.0 &               WISE &     55.8 &   \citet{wright2010} \\
    13.02 &           51.2$\pm$5.4 &                IRS &     44.2 &            this work \\
    14.87 &           44.7$\pm$4.6 &                IRS &     34.0 &            this work \\
    16.99 &           40.0$\pm$4.4 &                IRS &     26.1 &            this work \\
    19.02 &           43.1$\pm$5.0 &                IRS &     20.9 &            this work \\
    21.31 &           46.7$\pm$7.3 &                IRS &     16.7 &            this work \\
    22.09 &           51.6$\pm$3.3 &               WISE &     15.5 &   \citet{wright2010} \\
    23.67 &           45.6$\pm$2.0 &               MIPS &     13.5 &     \citet{chen2012} \\
    24.48 &           59.0$\pm$6.0 &                IRS &     12.6 &            this work \\
    27.45 &           77.3$\pm$7.6 &                IRS &     10.1 &            this work \\
    30.50 &          96.7$\pm$10.0 &                IRS &      8.1 &            this work \\
    33.55 &         137.9$\pm$18.6 &                IRS &      6.7 &            this work \\
    60.00 &         601.0$\pm$48.1 &               IRAS &      2.1 &       \citet{moshir} \\
    70.00$^{*}$ &         690.1$\pm$48.6 &               PACS &      1.5 &            this work \\
    71.42 &         655.0$\pm$44.4 &               MIPS &      1.5 &     \citet{chen2012} \\
   100.00$^{*}$ &         675.1$\pm$47.6 &               PACS &      0.7 &            this work \\
   160.00$^{*}$ &         462.4$\pm$32.7 &               PACS &     0.28 &            this work \\
   250.00 &         213.4$\pm$12.9 &              SPIRE &     0.12 &            this work \\
   350.00 &          120.3$\pm$8.7 &              SPIRE &     0.06 &            this work \\
   500.00 &          63.6$\pm$10.2 &              SPIRE &     0.03 &            this work \\
\enddata                                                                        
\tablenotetext{*}{The emission is spatially resolved at these wavelengths.}                                                                                 
\end{deluxetable*}   
The resulting SED with the stellar photosphere model
(Sect.~\ref{stellarprops}) is presented in Fig.~\ref{fig2}a. The IRS
spectrum shows that IR excess is present even at
$\sim$10${\micron}$. No silicate features are apparent, implying the
depletion of small grains. We assumed optically thin dust emission in
the modeling, and as a first simple model, we fitted the excess by a
single temperature modified blackbody, where the emissivity is equal
to 1 at $\lambda \leq \lambda_0$ and varies as
$(\lambda/\lambda_0)^{\rm -\beta}$ at $\lambda > \lambda_0$. Following
\citet{williams2006}, we adopted ${\lambda_0}$=100{\micron}. We used a
Levenberg-Marquardt algorithm in the modeling, and utilized an
iterative method to compute color corrections
\citep{moor2006}. Table~\ref{tab2} lists the best-fit parameters and
the reduced $\chi_r^2$ of the fitting. Fig.~\ref{fig2}a shows that the
best-fit model underestimates the observed excess, especially at
shorter wavelengths. This discrepancy can be explained if the emitting
dust grains -- similarly to many other debris systems
\citep[e.g.][]{morales2011} and our Solar System -- are distributed in
two spatially separated rings. With this assumption, we used a
two-component model, where grains in the warmer component act like
blackbodies, while the emission of the outer ring is described by the
modified blackbody as defined above. As Fig.~\ref{fig2}a and the
improved $\chi_r^2$ value (Table~\ref{tab2}) indicate, this model fits
the SED better over the whole studied wavelength range. The fractional
luminosity of the rings was computed as $f_{\rm dust} = {L_{\rm
    dust}}/{L_{\rm bol}}$. The radius of the dust ring(s) was
estimated using the following formula \citep{backman1993}:
\begin{equation}
\frac{r_{\rm dust}}{AU} = \left(\frac{L_{\rm bol}}{L_{\sun}}\right)^{0.5} \left(\frac{278\,K}{T_{\rm dust}}\right)^2.
\end{equation}
Because of the blackbody assumption, the resulting $r_{\rm dust}$
values {are lower limits}
The obtained
fundamental disk properties are listed in Table~\ref{tab2}.

The dust mass of the disk was estimated using the following formula:
\begin{equation}
M_{dust} = \frac{F_{\nu} d^2}{ B_{\nu}(T_{dust}) \kappa_{\nu}}, 
\end{equation}
where $F_{\nu}$ is the measured SPIRE flux at 500{\micron}, $d$ is the
distance to the source (90.4~pc), $\kappa_{\nu} = \kappa_{0}
(\frac{\nu}{\nu_0})^{\beta}$ is the mass absorption coefficient,
$B_{\nu} = \frac{2 \nu^2 k T_{dust}}{c^2}$ is the Planck function
using the Rayleigh-Jeans approximation. Assuming a $\kappa_{0} =
2$\,cm$^2$\,g$^{-1}$ at $\nu_0 = 345$~GHz \citep[e.g.][]{nilsson2010}
and taking $\beta = 0.41$ and $T_{dust} = 57$~K from Table~\ref{tab2}, we derived a dust
mass of 0.5$\pm$0.1~M$_\oplus$.

\begin{deluxetable}{lccc}
\tabletypesize{\scriptsize}
\tablecaption{Disk properties derived from SED fitting \label{tab2}}
\tablewidth{0pt}                                                                                                                                                                    
\tablecolumns{4}
\tablehead{\colhead{} & \colhead{Single temperature} & \multicolumn{2}{c}{Two temperature}  \\
\colhead{} & \colhead{} & \colhead{Warm comp.} & \colhead{Cold comp.} 
}
\startdata
$\chi_r^2$     & 4.6  &  \multicolumn{2}{c}{0.9} \\ 
$T_{\rm dust}$ [K] & 64$\pm$1  & 187$\pm$26  & 57$\pm$1.5  \\
$\beta$        & 0.17$\pm$0.06  & 0  &  0.41$\pm$0.07 \\
\hline
$r_{\rm dust}$ [AU] & 50$\pm$2.4  & 5.9$\pm$1.6  & 63.7$\pm$4.4  \\
$f_{\rm dust}$ [10$^{-3}$]   & 1.5$\pm$0.1  & 0.14$\pm$0.10  & 1.4$\pm$0.2  \\
\enddata   
\end{deluxetable} 

\subsection{CO data}

No CO emission was detected in our APEX spectrum. We converted antenna
temperatures to flux densities using conversion factors from the APEX
web
page\footnote{\url{http://www.apex-telescope.org/telescope/efficiency/}}
and computed an upper limit for the CO(3--2) line flux ($S_{CO(3-2)}$)
as $I_{\rm rms} \Delta{v} \sqrt{N}$, where $I_{\rm rms}$=0.95\,Jy is
the measured RMS noise at the systemic velocity of the star,
$\Delta{v}$=0.42\,km\,s$^{-1}$ is the velocity channel width, and $N$
is the number of velocity channels over an interval of
10\,km\,s$^{-1}$, which covers the expected line width. The
resulting upper limit is 1.95\,Jy\,km\,s$^{-1}$ for $S_{CO(3-2)}$.
Assuming optically thin emission and using a range of excitation
temperatures between 20~K and 60~K (the latter value corresponds to
the dust temperature in the outer ring, see Sect.~\ref{modelling}),
the upper limit for the total CO mass is between 1.4$\times$10$^{-4}$
and 1.7$\times$10$^{-4}$~M$_\oplus$.

\section{Discussion}
\label{discussion}

\begin{figure*} 
\includegraphics[scale=.45,angle=0]{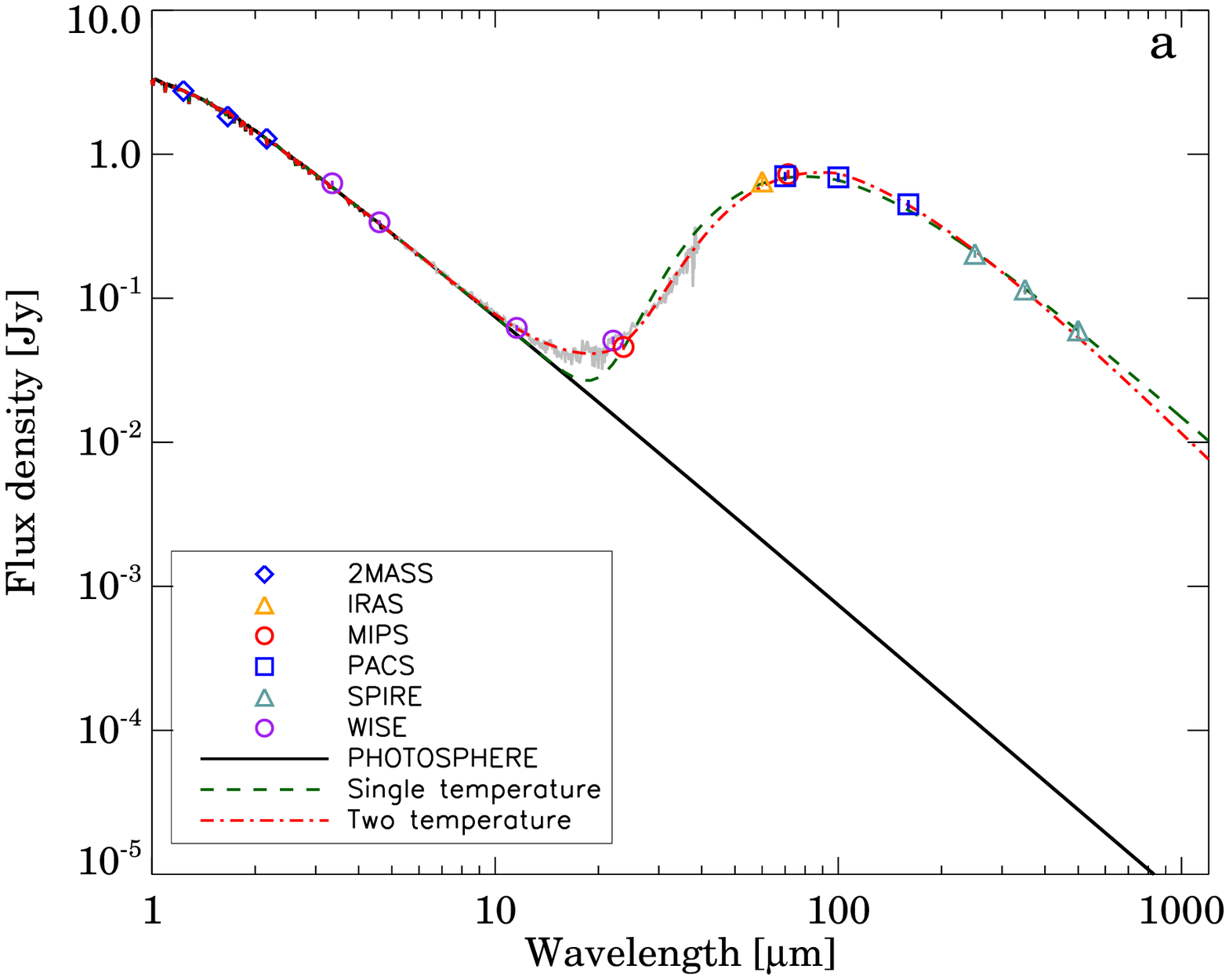}
\includegraphics[scale=.45,angle=0]{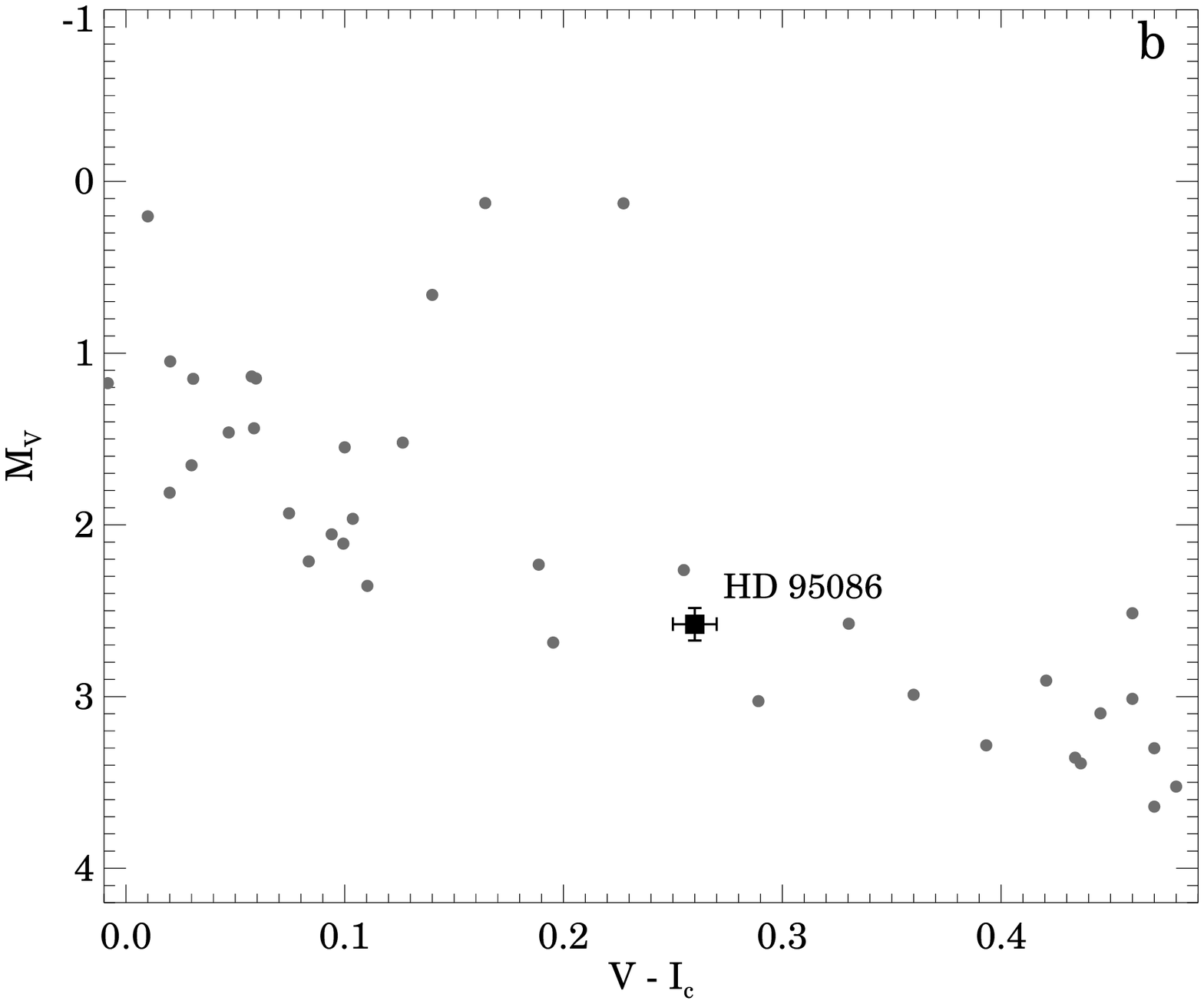}
\caption{ {\sl Left: Color-corrected SED of the source {and} fitted models. 
Right: Absolute V band magnitude versus V$-$I$_{\rm c}$ 
color diagram for known members of LCC with parallax error $<$10\% (gray dots) and for HD95086 (black square). 
Photometric data were taken from the {\sl Hipparcos} catalog and dereddened 
using extinction values {from} \citet{chen2011,chen2012}.}
\label{fig2}
}
\end{figure*}

Analyzing {\sl Hipparcos} astrometric data, \citet{dezeeuw1999}
identified a total of 180 probable members in the LCC
association. Among their proposed members, HD95086 has the lowest
membership probability of 41\%. In order to re-check the membership
status of HD95086, we used our new radial velocity and {\sl Hipparcos} astrometric
data to compute a Galactic space
motion of U\,=\,$-$10.5$\pm$1.0, V\,=\,$-$22.3$\pm$1.8,
W\,=\,$-$5.2$\pm$0.4\,kms$^{-1}$. According to \citet{chen2011},
the characteristic space motion of LCC is U\,=\,$-$7.8$\pm$0.5,
V\,=\,$-$20.7$\pm$0.6, W\,=\,$-$6.0$\pm$0.3\,kms$^{-1}$, thus there is
a good kinematic match between our target and LCC. The position of
HD95086 in the color-magnitude diagram in Fig.~\ref{fig2}b fits well
to the locus defined by LCC members, indicating that its isochrone age
is consistent with that of the cluster. Based on these findings, we
propose that HD95086 is a very probable member of LCC. With its
distance of 90.4~pc, HD95086 is one of the closest member of the
association.

\citet{pecaut2012} derived an age of 17~Myr for LCC based on isochrone
fitting while \citet{song2012} proposed 10~Myr from lithium content of
late-type LCC stars. The latter estimate is close to the timescale
on which primordial gas is believed to disappear
\citep[e.g.][]{mamajek2009}. Thus, if the 10 Myr age estimate proves
to be true, we cannot exclude that this disk would contain substantial
amount of primordial gas similarly to transitional disks.  Converting
our upper limit on the CO mass to total gas mass assuming a canonical
H$_2$/CO abundance ratio of 10$^4$, the estimated gas mass in the disk
of HD95086 is $\lesssim$0.12~M$_\oplus$ ($\lesssim$3.7$\times
10^{-6}$~M$_\odot$).  This is several orders of magnitude below the
typical Herbig~Ae and transitional disk masses.  This result and the
low gas-to-dust mass ratio of $\lesssim$0.25 implies that HD95086
harbors a gas-poor debris disk.

The debris disk of HD95086 has a high fractional luminosity of
1.5$\times$10$^{-3}$. With this value, it belongs to the top ten
highest fractional luminosity debris systems in the {solar neighbourhood} \citep[d
  $<$ 120~pc,][]{moor2006,rhee2007}. Interestingly, these systems have
other common properties: they are younger than 100~Myr, and most of
them belong to different young moving groups and associations.
We found that the HD95086 disk contains two dust belts: a warm one
with a characteristic temperature of 187~K, and a colder one with
57~K. Such structures seem to be relatively
common. \citet{morales2011} found that a significant fraction of
debris disks {harbors} two spatially distinct dust components. The
characteristic dust temperatures for the inner and outer dust belts (190~K and 60~K)
were found to be very similar independently of the host stars'
luminosity, implying that the formation of dust belts is controlled by
a temperature-sensitive mechanism. Several other members of the young
massive debris disk group also share similar properties
\citep[e.g.,][]{su2009,roberge2013,donaldson2013}. Because the
characteristic temperature of the warm component is slightly above the
ice evaporation temperature, \citet{morales2011} proposed that the
formation of the warm component is related to sublimation of icy
planetesimals crossing the snow line, or due to collisions in an
asteroid belt-like system formed just interior to the snow line
\citep[see also][]{martin2013}. One of these mechanisms could also
work in the case of HD95086.

To determine the relative location of the two dust belts and the
planet candidate, we need to constrain the inclination of the
planetary orbit. From our resolved Herschel observations, the disk
inclination is $\sim$25{\degr}. The star's equatorial plane is
probably not very different from this, because the observed $v \sin i$
= 20\,km\,s$^{-1}$ is unusually low compared to stars with similar
masses \citep{zorec2012}, implying a low {stellar} inclination. 
Assuming that the disk and the planet's orbit are coplanar, and
taking the projected orbital radius of 56\,AU from \citet{rameau2013},
the deprojected orbital radius would be 62\,AU. We derived a radius of
64\,AU for the outer dust belt. Since we adopted blackbody grains,
this is a lower limit, because smaller grains with the same
temperature can be located farther from the star. Thus, it is possible
that the planet candidate is situated just inside the cold
outer dust belt. Further monitoring of the planet candidate will
reveal whether it orbits within the outer dust belt. 
If it stays inside, it might sculpt the inner edge
of the belt, and may induce azimuthal asymmetries in the dust
distribution. We note that the planetary system of HD95086 resembles
the one around HR8799. Both systems harbor a warm inner dust belt
and a broad colder outer disk and giant planet(s) between the two
dusty regions.

The short lived dust grains in debris disks are believed to be
replenished by collisions between larger bodies. For destructive
collisions, the collision velocity must exceed a critical value and it
requires a dynamical perturbing force. According to theory, this
perturbation is linked to 1000~km-sized planetesimals formed within
the planetesimal ring \citep[self-stirring,][]{kb2008}, or the
presence of planet(s) in the system \citep[planetary
  stirring,][]{mustill2009}. The formation of 1000~km-sized
planetesimals is a slow process at large orbital radii.  Adopting an
age of 17~Myr for the system and using Eqs.~27,41 from \citet{kb2008},
we found that, even in a disk with an unusually high initial surface
density (ten times higher than the minimum-mass solar nebula), the
formation of such large bodies is limited to $\lesssim$70~AU. Our PACS
images show that the disk around HD95086 is spatially extended with a
diameter of 540~AU. While large dust grains have the same spatial
distribution as the parent planetesimals, the distribution of small
grains \citep[smaller than the blowout limit of $\sim$1.8{\micron}, estimated following][]{wyatt2008}
can be more extended, because
they are expelled by stellar radiation pressure and form a dust
halo. However, the contribution of such small grains to the
100 and 160{\micron} {flux} is small, suggesting that HD95086 harbors an
extended planetesimal belt whose dynamical stirring cannot be
explained by self-stirring. \citet{mustill2009} claim that
planetary stirring can be faster than self-stirring in the outer
regions of planetary systems. Using their formulae, we found that the
candidate planet can dynamically excite the motion of planetesimals
even at 270~AU via their secular perturbation if its orbital
eccentricity is larger than $\approx$0.4. Alternatively, without
effective stirring, collisions with low velocity can lead to
merging grains, thus we cannot exclude that the outer disk region
contains primordial dust as well.

\acknowledgments
We thank our anonymous referee whose comments improved
the manuscript.
This project was supported by the Hungarian OTKA grants K101393 and
K104607, the PECS-98073 program of the European Space Agency (ESA) and
the Bolyai Research Fellowship of the Hungarian Academy of Sciences.

{\it Facilities:} \facility{Herschel}, \facility{APEX}, \facility{Spitzer}.


\end{document}